\documentstyle[twoside,fleqn,espcrc2]{article}

\def\psibar{\overline{\psi}}

\def\Square{{\vbox
{\hrule height 0.6pt\hbox{\vrule width 0.6pt\hskip 3pt

\vbox{\vskip 6pt}\hskip 3pt \vrule width 0.6pt}\hrule height 0.6pt}}}
\def\Dsl{\hbox{/\kern-.6700em\it D}} 
\def\hf{\frac{1}{2}}
\def\veps{\varepsilon}

\def\Scl{{\cal L}}
\def\Scd{{\cal D}}

\def\L{\Lambda}

\def\d{{\rm d}}
\def\del{\partial}
\def\delbar{\overline{\partial}}
\def\inv{^{-1}}
\def\dsl{\hbox{/\kern-.5300em$\partial$}}
\def\IR{\relax{\rm I\kern-.18em R}}
\font\cmss=cmss10 \font\cmsss=cmss10 at 7pt
\def\IZ{\relax\ifmmode\mathchoice
{\hbox{\cmss Z\kern-.4em Z}}{\hbox{\cmss Z\kern-.4em Z}}
{\lower.9pt\hbox{\cmsss Z\kern-.4em Z}}
{\lower1.2pt\hbox{\cmsss Z\kern-.4em Z}}\else{\cmss Z\kern-.4em Z}\fi}
\def\M{\Lambda}

\def\delbar{\overline\del}
\def\inbar{\vrule height1.5ex width.4pt depth0pt}
\def\IC{\relax\thinspace\hbox{$\inbar\kern-.3em{\rm C}$}}

\newcommand{\AmS}{{\protect\the\textfont2
  A\kern-.1667em\lower.5ex\hbox{M}\kern-.125emS}}

\title{Duality and Global Symmetries}

\author{Fernando Quevedo\address{Instituto de F\'{\i}sica,\\
Universidad Aut\'onoma de M\'exico\\
Apartado Postal 20-364, 01000\\
M\'exico D.F., M\'exico\\
Email: fernando@ft.ifisicacu.unam.mx
}%
        \thanks{Lectures given at 33rd Karpacz Winter School, February 1997.
Preprint IFUNAM FT97-07, hep-th/9706210}}
        
\begin{document}

\begin{abstract}
Duality symmetries are reviewed. A sufficient condition
for duality is the existence of a global symmetry.
This can be used as a guideline to systematically prove
duality between different field theories.
 Bosonization and $T$- duality, for abelian
and non-abelian global symmetries, are discussed in detail as well
as duality for general antisymmetric tensor field theories.
In this case, the presence of topological defects  break the global 
symmetry but duality survives manifesting itself in a 
different way. Open questions and current limitations of this approach
to prove all known dualities are discussed.
\vspace{12pt}

\begin{center}
{ \em To the hundreds of thousands
 of Guatemalan citizens who 
sacrificed their lives  during more than }
{\em three decades of civil  war,  especially to those I had the privilege 
 to know and love, and who I now miss.}
\end{center}
\end{abstract}

\maketitle

\section{INTRODUCTION}
The current excitement about duality in field and string theories
leads us  to believe that duality is indeed an exact symmetry of
many  physical systems. But in many cases we can only
rely on indirect evidence for duality rather than proving it 
explicitly.  We would like to know precisely
when two theories can actually be shown to be dual to each other. 
The general answer to this question is unknown but 
a sufficient condition for the existence of duality is the existence of 
a global symmetry in the original theory. Given any field theory
with a global symmetry, there is a well defined prescription to 
find the dual theory. This  brings us a long way towards explicitly showing 
the equivalence between two different theories and puts many dualities 
on an equal setting. In particular we will see that bosonization can
be proven exactly in the same way as we prove $T$ duality.

First  I would like to remark  that, before the recent 
developments,   duality had  a long 
history. It was  known to Maxwell 
that his equations in vacuum were invariant under the 
exchange of electric and magnetic fields, but it was until 1931 that
Dirac used this duality to introduce the magnetic monopole in order 
to keep electric/magnetic duality valid  in the presence of sources.
Independent of this, Kramers and Wannier discovered temperature  duality in the
2D Ising model allowing them to learn about the phase structure of the
model before Onsager's solution \cite{savit}. This was then generalized to several
lattice and continous models leading to the Higgs/confinement 
duality of 't Hooft and Mandelstam as well as the Montonen-Olive
conjecture for duality in 4D gauge theories over two decades ago  \cite{olive}.
At about the same  time, independent to this development,
 Luther and Preschel realized that  
fermions and bosons in 2D systems are equivalent  \cite{luther}.
Bosonization  has lead to many applications
in condensed matter physics and high energy  theory.
All these `dualities' had in common that they  relate strong to 
weak coupling (or temperature) and that they exchange elementary
states by topological defects or solitons (electric/magnetic charges)
although they were discovered  independently and using different techniques.

During the past decade, Witten introduced non-abelian 
bosonization \cite{witten1},  hidden symmetries of supergravity theories
were discovered containing duality  \cite{sugra}\ and,   in string theory,
the subject started  with the realization of 
$R\leftrightarrow 1/R$ duality in toroidal compactifications  \cite{ponjas},
  \cite{giveon}. 
Later on, this duality was  understood in terms of the standard duality
of 2D  non-linear sigma models studied in supergravity  \cite{buscher}
, \cite{rover}\
  and
it was extended to
more general backgrounds, 
including backgrounds with non-abelian isometries  \cite{nonab}. Duality
originating from symmetries of the string 2D sigma model is 
presently known as  $T$-duality. These developments lead 
also to the conjecture of a possible duality in condensed matter
systems such as the quantum Hall effects, almost ten years ago  \cite{shapere}.
This has acquired more relevance recently  after the
spectacular experimental evidence found last year  \cite{exper}.

The subject of this school has its origin on the more recent developments
related to strong/weak coupling or $S$-duality in supersymmetric
field and string theories which started during the present
 decade
and are the main  subject  of research  at present  \cite{sdual}, \cite{sdualrev}.
Most of the recent developments will be reviewed by the other
speakers. I will concentrate on the concrete question: under which
circumstances
can we actually prove duality between two different theories?.
In string theory the methods presented here apply mostly to
$T$-duality. For $S$-duality we cannot prove it explicitly because
at present
there is no non-perturbative formulation of string theory,
that is why we have to rely  on   indirect evidence for this duality.
However,
one of these evidences is the duality between the two effective
field theories of the massless modes of both (dual) strings. That
corresponds  to duality of antisymmetric tensor field theories
which  we will discuss in these lectures.

The aim of these lectures is then to review the explicit prescriptions  for
dualization. If there is any field theory with a global symmetry
then we can dualize it by gauging the corresponding symmetry and
introducing a Lagrange multiplier constraint that sets the gauge
field strength to zero  \cite{rover}.
 Depending on the nature of the symmetry we can have
abelian or non-abelian duality and depending on the nature
of the original degrees of freedom we can have duality for scalar fields
in 2D ($T$-duality), for electromagnetism or arbitrary antisymmetric
tensors ($S$-duality), or duality for fermions $i.e.$ bosonization.

The lectures are organized as follows.
In section 2 we  discuss abelian duality for two
specific systems: arbitrary antisymmetric tensors in any
dimension and fermions in two and higher dimensions. The particular 
case of rank zero tensors in 2D is discussed in some detail
($T$-duality) and examples, such as toroidal compactifications and 2D 
black holes, are presented. In section 3 duality is extended to the
nonabelian case. Here we restrict ourselves to non-abelian $T$-duality
and non-abelian bosonization. Some 
proposed generalizations of non-abelian duality are also presented.
Finally in section 3 we discuss duality for the systems of
section 1 when the presence of topological defects, such as magnetic monopoles,
break the original global symmetry. For pedagogical reasons,
we illustrate the general
case by studying mostly  QED. First we explictly show duality 
in the presence of magnetic monopoles. Then, after these monopoles
condense, we present the effective Lagrangian describing 
 the so-called `confining string' description 
of the confining phase of the theory, we also prove that duality is 
manifest even in this case. We close the section with 
the examples of $T$-duality after vortex condensation for the bosonic
string case and describe how duality is not broken by the
Kosterlitz-Thouless phase transition. 
We close the lectures mentioning some open questions for the
techniques presented here and some speculations.

\section{ABELIAN DUALITY}

We will review in this section duality in systems with abelian
global symmetries. First we discuss the 2D case of the string
worldsheet action. Then we generalize it to higher dimensions
and fermionic systems.

\subsection{Abelian $T$-Duality}

We will briefly review here the standard duality corresponding 
to string  backgrounds with abelian
isometries.
 The worldsheet action for the bosonic string is:
\begin{eqnarray}
S&=&\frac{1}{4\pi\alpha'}\int \d^2z\,
\left(\frac{}{}\ Q_{MN}(X_{Q})\,\del X^{M} \delbar X^{N}\right.\nonumber\\
& &+
 \frac{\alpha'}{2} \left. R^{(2)}\Phi(X_Q)\ \right)
\end{eqnarray}
 where $Q_{MN}\equiv G_{MN}+B_{MN}$. $G_{MN}$ and $B_{MN}$
being the metric and antisymmetric tensor of the target space theory
and $\Phi$ is the dilaton field. They are `couplings' in the 2D
theory.  $X^M$ are  scalar fields
of the 2D theory  which are the  coordinates
in target space. $ R^{(2)}$ is the scalar curvature in 2D.
We take the target space to have the critical dimension,
$D=26$ for purely bosonic strings and $D=10$ for supersymmetric
strings, in that case the action above would represent only
(part of) the bosonic sector of the theory.
 In a
background with $N$ commuting isometries, this can be written as

\begin{eqnarray}
S& = &\frac{1}{4\pi\alpha'}\int \d^2z\,
\left(\ \frac{}{} Q_{\mu\nu}(X_{\alpha})\,\del X^{\mu} \delbar X^{\nu}\
\right.  \\ & & \left. + Q_{\mu n}(X_{\alpha})\del X^{\mu} \delbar
X^n
 + Q_{n \mu}(X_\alpha)\del X^n \delbar X^{\mu}\right.\nonumber  \\
& & \left.  + Q_{mn}(X_\alpha)\del X^m \delbar X^n +
  \frac{\alpha'}{2} R^{(2)}\Phi(X_\alpha)\right)\ .\nonumber
\end{eqnarray}
Lower case
latin indices $m,n$ label the isometry directions. Since the
action  depends on the $X^m$ variables
only through their derivatives
(there exists  the abelian  global symmetry
 $X^m\leftrightarrow X^m +$ constant), we can write it in
first order form by introducing
variables $A^m$ and adding an extra term to the action
\hbox{$\L_m(\del \overline A^m - \delbar A^m)$} which imposes
the constraint $A^m=\del X^m$.
Integrating over the Lagrange multipliers $\L_m$ returns
us to the original action (1). On the other hand
performing partial integration and solving for
$A^m$ and $\overline A^m$, we find the dual action $\tilde S$
which has an identical form to $S$ but with
the dual background given by
 \cite{buscher}, \cite{ginsparg} 
\begin{eqnarray}
\tilde Q_{mn}& = &(Q\inv)_{mn}\\
\tilde  Q_{\mu \nu}& =& Q_{\mu \nu} - Q_{\mu m}\,(Q\inv)^{mn}\,Q_{n\nu}\nonumber\\
\tilde  Q_{n \mu}& = &(Q\inv)_n^{\ m}\, Q_{m \mu}\nonumber\\
\tilde  Q_{\mu n}& =& - Q_{\mu m}\,(Q\inv)^m_{\ n}\ .\nonumber
\end{eqnarray}
To preserve conformal invariance, it can be seen 
that the dilaton field has to transform as
$\tilde \Phi = \Phi - \log\det G_{mn}$.
Equations (2) reduce to the usual
duality transformations for the toroidal compactifications 
 in the case $Q_{m\mu}=Q_{\mu m}=0$ and can map
 a space with no
torsion ($Q_{m\mu} =Q_{\mu m}$)  to a space
with torsion ($\tilde Q_{m\mu}=-\tilde Q_{\mu m}$).

Duality then states that the two dual theories defined by 
the actions $S[Q,\Phi]$ and $\tilde S[\tilde Q,\tilde \Phi]$ give the same
partition function $Z$ and therefore they are
physically equivalent. A more precise statement could be
made by coupling the system to external sources so that 
the partition function and its derivatives with respect to
the source are equivalent, meaning that also all the
amplitudes are equivalent after a duality transformation.
 
An equivalent interpretation of the duality process
just described is given by gauging the symmetry,
replacing
$\del X^m$ with $D X^m = \del X^m + A^m$ and the
term $\int d^2z\ \L_m(\del \overline A^m - \delbar A^m)$
is added to the action.  This extra term
imposes the vanishing of the field strength $F$
of the gauge fields after integration over the
Lagrange multipliers $\L_m$.
This implies that locally the gauge field must be pure gauge,
$A^m = \del X^m$.
The gauge fixing can be done either by
choosing the gauge fields to vanish or by taking
$X^m=0$ (a unitary gauge). In both cases this
reproduces the original action.
The dual theory is obtained by instead integrating out the
gauge fields and then fixing the gauge.

This is the prescription that can be systematically generalized:
whenever  there is a global symmetry we can gauge it and impose
the corresponding field strength to zero. Changing the order of
integrations we end up either with the original theory or with the
dual theory where the Lagrange multiplier becomes the fundamental
dual variable.

As an illustrative  case we will
 discuss the circle compactifiction in more detail. In that 
case there is one isometry given by the shift on the circular coordinate.
The corresponding metric component is just $R^2$, the radius of the circle.
Therefore the duality equations above reduce to $R\leftrightarrow \alpha'/R$
 duality.
Furthermore, being this a free 2D field theory, canonical quantization 
is straightforward, giving rise to a mass formula:
\begin{equation}\alpha'\, M^2={n^2}\,
 {\alpha'\over R^2}+m^2\, {R^2\over \alpha'}
+2\, \left(N_L+N_R-2\right)
\end{equation}
The  physical states are labeled by the integers $m,n$ where $P=n/R$
are the quatized momenta in the compact direction whereas $W=m R$
are the solitonic winding states with $m$ the number of times the string wraps
around the circle. We can immediately observe that the mass formula is invariant 
under the duality transformation $R\leftrightarrow \alpha'/R$ as long as the roles 
of the winding and momentum states are also exchanged. This is an example of
a generic feature about duality that exchanges elementary with solitonic states. 

This relation can be made more explicit  by absorbing the
 radius $R$ in the definition 
of the coordinate $X$ which then has periodicity $X\equiv X+2\pi R$.
In this situation we denote the dual coordinate by $\tilde{X}$.
Probing the theory with external sources $J,\overline J$ the original
action is
\begin{equation}
S=\frac{1}{\pi\alpha'}\int d^2z \left(\del X\delbar X+J\delbar X+\overline J
\del X+\cdots\right)
\end{equation}
whereas the dual action is
\begin{equation} 
\tilde S=\frac{1}{\pi\alpha'}\int d^2z \left(\del \tilde X\delbar \tilde X-J\delbar
 \tilde X+\overline J
\del \tilde X+ J\overline J\right)
\end{equation}
Notice the difference   in the couplings to the currents and  the
presence of the contact term $ J\overline J$ in the dual action
\footnote{This term  is precisely what is needed in order that both dual theories
agree after taking care of time ordering in the corresponding amplitudes
(I thank C.P. Burgess for explaining this to me.)}.
By comparing the coefficients of the currents we can see the
relations $\del X\leftrightarrow\del \tilde X$ and 
$\delbar X\leftrightarrow - \delbar\tilde X$.
This can be deduced by using the field equations
for both $A,\overline A$ and $\L\equiv\tilde X$ in the first order
action and only means that
(using 2D  cartesian coordinates rather than complex  ones)
 the two dual variables are related by
the standard Poincar\'e duality of forms,
 $\del_\alpha\tilde X=\veps_{\alpha\beta}
\del^\beta X$. Which in turn implies that the 
field equations $\del_\alpha\del^\alpha X=0$
transform into Bianchi identities $\veps_{\alpha\beta}
\del^\alpha\del^\beta\tilde X$ and viceversa.

This information,  combined with the wave equation
$\del\delbar X=\del\delbar \tilde X=0$ implies that $X(z,\overline z)=
X_L(z)+X_R(\overline z)$ and $\tilde X(z,\overline z)=
X_L(z)-X_R(\overline z)$. Comparing the zero-modes in the 
oscillator expansion of both variables:
\begin{eqnarray}
X&=&P\tau+W\sigma +\cdots\\
& =& \hf p_L\ln z+\hf p_R\ln\overline{z}+\cdots\nonumber
\end{eqnarray}
where $p_{L,R}\equiv P\pm W$ are the left and right-moving momenta
and the ellipsis refer to the oscillator terms.
The dual variable is then 
\begin{equation}
\tilde X=\hf p_L\ln z-\hf p_R\ln\overline{z}+\cdots
\end{equation}
 From this we can see that the two dual theories are
related by the exchange $P\leftrightarrow W$, also that 
the hamiltonian (and so the mass formula) depends only
on the combination $p_L^2+p_R^2$ which is duality invariant.
Since the periodicity of $X$ gave rise to the quantization of the
momenta $P$ and the momenta  of the dual variable are given by $W$
we can then easily extract the periodicity of $\tilde X$ to be
$\tilde X\equiv \tilde X+2\pi\alpha'/R$ (an overall integer factor
$p$ can be considered, coinciding with the Dirac quantum
\cite{carlome}, but 
modular invariance of the string theory restricts to $p=1$).

The simplest generalization of this system is compactification on a 2D
torus instead of a 1D circle. This extension is straightforward but
very interesting. Duality gives the transformation of the 2D
matrix  $G_{ij}+B_{ij}\leftrightarrow
(G_{ij}+B_{ij})^{-1}$. But now, since the purely antisymmetric tensor
part of the action (1) is topological, there is also a symmetry
$B\rightarrow B+$ integer, where $B$ is defined from $B_{ij}=B\veps_{ij}$.
Since this symmetry does not commute with duality, they together
generate an infinite dimensional discrete symmetry. This is better
expressed in terms of the complex parameters or {\it moduli}:
\begin{eqnarray}
T&=& B+i\, \sqrt{G}\\
U&=&\frac{G_{12}}{G_{22}}+i\, \frac{\sqrt{G}}{G_{22}}\nonumber
\end{eqnarray}
$U$ is the standard modular parameter of any geometrical 2D torus and
it is usually identified as the `complex structure' modulus. $T$ is the
`K\"ahler structure' modulus (since the torus is a complex K\"ahler
manifold) and its imaginary part measures the overall size of the torus,
since $G$ is the determinant of the 2D metric. It then plays the same
role as the radius $R$ for the 1D circle.

In terms of $T$ and $U$ we can write the left- and
right-moving momenta as:
\begin{eqnarray}
p_L^2=\frac{1}{2U_2T_2}\|(n_1-n_2\, U)-T\, (m_2+m_1\, U)\|^2\nonumber\\
p_R^2=\frac{1}{2U_2T_2}\|(n_1-n_2\, U)-\overline{T}(m_2+m_1\, U)\|^2
\label{momenta}
\end{eqnarray}
The mass formula, depending on $p_L^2+p_R^2$, shows the following symmetries:
\begin{equation}
U\rightarrow\frac{a\, U+b}{c\, U+d}\qquad T\rightarrow\frac{a\, T+b}{c\, T+d}
\qquad T\leftrightarrow U.
\end{equation}
Where $a,b,c,d$ are integers satisfying $ad-bc=1$.
The first transformation is the standard 
$SL(2,\IZ)_U$ `modular' symmetry of 2D
tori and is independent of string theory; it is purely geometric.
The second transformation is a stringy $SL(2,\IZ)_T$ named 
$T$-duality and it is a generalization of $R\leftrightarrow 1/R$
for the 2D case. Again this is a symmetry as long as we also transform
 momenta $m_1,m_2$ with winding $n_1,n_2$.
The field $T$ is then at the origin of the name $T$- duality (but the 
field $U$ is unrelated with the so-called $U$-duality).
 The third symmetry exchanges the
complex structure $U$ with the K\"ahler structure $T$ and
it is called `mirror symmetry'.
The moduli $U$ and $T$ each parametrize a complex plane
$SL(2,\IR)/O(2)$, the duality symmetry implies that they can only live
in the fundamental domain defined by all the points of the product of
 complex spaces
 $SL(2,\IR)/O(2)\otimes SL(2,\IR)/O(2)\cong O(2,2,\IR)/(O(2)\times O(2))$
 identified under the 
duality group $SL(2,\IZ)_U\times SL(2,\IZ)_T
=O(2,2,\IZ)$.

This is the situation that gets generalized to higher dimensions.
In general, compactification on a $d$-dimensional torus has the
Narain \cite{narain}\ moduli space ${\cal M}=O(d,d,\IR)/O(d)\times O(d)$ with 
points identified under the duality group $O(d,d,\IZ)$.
For the heterotic string with 16 extra left moving coordinates
${\cal M}=O(d+16,d,\IR)/O(d+16)\times O(d)$ with a similar 
modification to the duality group.

A slightly more complicated example is the 2D black hole defined by the metric
$-du\, dv/(1-uv)$\cite{witten2}.
Diagonalizing the metric, it can be easily seen that the asymptotically 
flat region $uv<0$ is defined by the metric
$dr^2-\tanh^2r\, dt^2$, the region between the horizon and the singularity
$0<uv<1$ has the metric $-dt^2+\tan^2t\, dr^2$ whereas beyond the
singularity, $uv>1$ it is $dr^2-\coth^2 r\, dt^2$. The geometry is
selfdual, but  the asymptotically flat region and the region beyond the
singularity are mapped to each other (global symmetry,
$t\rightarrow t+$ constant implies $\tilde{G_{tt}}=1/G_{tt}$).
Whereas the region between the singularity and horizon is selfdual,
since $\cot t=\tan(t-\pi/2)$.
This is remarkable because, in particular, duality is exchanging 
singularity$\leftrightarrow$horizon
\cite{giveon2}. Moreover, in simple
extensions of this geometry to 3D, the geometry is a {\it black string}
(in the sense that the singularity is not point like but one-dimensional)
\cite{ginsparg2}, \cite{ginsparg}\
which is no longer selfdual. It is dual to another black string where now the 
singularity of one geometry is mapped to a regular surface in the
asymptotically flat region  of the other geometry and on which,
contrary to the horizon, classical computations are dominant. This raises
the hope that singularities may be consistently treated in string theory by
performing calculations in the dual geometry.

It is then interesting to investigate if these properties hold for
physically more interesting objects such as 4D black holes.  
 In
particular the Schwarzschild
metric
\begin{eqnarray}
ds^2=-(1-{2M/r}) dt^2 + \frac{dr^2}{1-{2M/r}}
               + r^2 d\Omega
\end{eqnarray}
times any CFT with $c= 22$ is a classical solution of bosonic 
string theory
to leading order in $\alpha'$.
Here $d\Omega=(d\theta^2 + \sin^2\theta d\phi^2)$ is the element
of solid angle. 
Direct application of the standard duality
transformation to this metric  for time
translations, gives the dual metric
\begin{eqnarray}
d\tilde s^2=-\frac{dt^2}{1-{2M/r}} + \frac{dr^2}{1-{2M/r}}
               + r^2 d\Omega
\end{eqnarray}
with the dilaton field now given by $\tilde\Phi=\Phi - \log(1-{2M/r})$.
This metric defines a geometry with naked singularities at
$r=0$ and $r=2M$, as it can be verified by
computing the curvature scalar
$R={4M^2\over (2M-r)r^3}$. It is easy to check
that the  string background equations    are satisfied by
the dual metric and dilaton $\tilde\Phi$. We have then
found a spherically symmetric solution of the
string background equations, which is not a black hole, but has naked
singularities  and
is dual to the Schwarzschild solution.

A similar analysis can be done for the 4D charged
dilatonic black hole. In this case the metric is
\cite{gibbons}
\begin{eqnarray}
ds^2 &=
     &-\frac{(1-{2M/r})}{(1-Q^2/Mr)}\,  dt^2  \\
& & + \frac{dr^2}{(1-{2M/r})
                   (1-Q^2/Mr)}
                   + r^2 d\Omega\ ,\nonumber
\end{eqnarray}
the dilaton field is $\Phi = -\log (1-Q^2/Mr)$ and the electric field
$F_{tr}= e^{\Phi}Q/(2r^2)$. It is very interesting
to note that the dual of this
solution with respect to time translations gives exactly the
same solution except that the mass parameter $M$ changes into
$ Q^2/2M$, therefore it relates
the black hole domain $Q^2<2M^2$ to
the naked singularity domain $Q^2>2M^2$.
In particular, the extremal solution $Q^2=2M$
is selfdual. 

Extracting physical implications to these results in 4D 
is premature because these solutions are only classical 
solutions of the leading order
 string background equations and the physically interesting regions
where the horizons and singularities are, usually correspond to strong 
coupling string theory where perturbative calculations are not
relevant. For the exact solutions not that much can be said
but  it can be shown that
$T$-duality associated with time translations in a
static spherically symmetric string background, exchanges the
ADM mass with the dilaton charge and the axion charge with the 
Taub-NUT parameter. Furthermore, as it will be seen in other lectures at this
conference, the existence of $S$-duality allows to understand
better the strong coupling regions of these geometries, leading to
the impressive calculation of the black-hole entropy/area relation
from the counting of microscopic states.

Before finishing this subsection  it is worth remarking some general
properties that have been found for abelian $T$-duality.
First it was shown by Ro\v cek and Verlinde that if the coordinate associated
with the abelian isometry is periodic, then the two dual theories 
define the same conformal field theory, if the coordinate is not 
periodic it is not established (for a discussion of this issue
see for example \cite{alvarez}\  and \cite{rob}\ ).
 Also in \cite{rover} and \cite{alvarez},
 global issues of abelian
$T$ duality where discussed, in particular by considering the 2D surface 
to be a torus, it  can be explicitly seen that the periodicity of the dual variable
coincides with the periodicity of the original coordinate and
duality survives at any order in string perturbation theory.
Furthermore, it was shown that $T$-duality is a remnant of a gauge symmetry
\cite{huet}\ 
in the sense that at the self dual point there is an enhanced gauge symmetry
since the Kaluza-Klein $U(1)$ symmetry of circle compactification
gets enhanced to $SU(2)$ at the selfdual radius $R=\sqrt{\alpha'}$
(take $m=n=\pm 1$ and $N_R=1$ in equation (4)
showing the existence of extra massless bosons which complete $SU(2)$).
 This symmetry
 is broken at any other point leaving duality as the discrete 
unbroken symmetry. This has been used to argue that $T$- duality
should also survive nonperturbative string effects.
The $O(d,d)$ symmetries of toroidal compactifications can 
be generalized to any background with $d$ abelian isometries
\cite{odd}.
Finally the explicit form of the duality transformation given above may be
corrected by considering higher order terms in the sigma model 
expansion parameter $\alpha'$, see for instance \cite{loops}. 
This has raised the interesting question of combining duality 
with the renormalization group flow in the sense that  the beta
function may transform in  a well defined way under duality
and could give information about the phase structure of the theory.

\subsection{General Antisymmetric Tensors}

The discussion of abelian $T$-duality can be immediately generalized
to arbitrary antisymmetric tensors of rank $h-1$, $B_{h-1}$ in $D$ dimensions
\cite{kalbramond}.
In this case the simplest action is
\begin{equation}
S=\int\, d^D x\left(\frac{1}{g^2}\, \left(\partial_{M_1}B_{M_2\cdots M_h}
\right)^2+
\cdots\right)
\end{equation}
The action is invariant under the shift of $A$ by an arbitrary closed form.
$\delta B_{h-1}=d\, \Omega_{h-1}+K_h$ with $K_h$ an arbitrary harmonic form.
The symmetry here is more general than the 2D case because it
includes a local symmetry defined  by $d\, \Omega$ and a global symmetry 
given by the shift in  $K$. This is the global symmetry we need for dualization.
Notice that gauging this symmetry involves a gauge field which is
an antisymmetric tensor of rank $h$, $A_h$.
Again we can dualize in two equivalent ways.  First we
construct a first order action  by substituting 
$d\, B$ in the action by $A_h$ and imposing the constraint
$dA_h=0$ by adding a Lagrange multiplier $\L_{D-h-1}$.
\begin{equation}
S_G=\int d^Dx\left(\frac{1}{g^2}\, (A_h)^2+\L_{D-h-1}\cdot(d A_h)\right).
\end{equation}
Where for simplicity we have defined
 $\L_{D-h-1}\cdot(d  A_h)=\epsilon^{M_1\cdots M_D}\partial_{M_2}
A_{M_2\cdots\M_{h+1}}\L_{M_{h+2}\cdots M_D}$.
Integrating $\L_{D-h-1}$ reproduces the original action whereas integrating 
$A_h$ leads to the dual action 
\begin{equation}
S=\int d^Dx\, \left(g^2(\del_{M_1} \L_{M_2\cdots\M_{D-h}})^2+\cdots 
\right).
\end{equation}
This is the higher dimensional generalization of the
$R\leftrightarrow 1/R$ duality of strings on a circle.
To make the analogy complete we are considering
{\it compact} antisymmetric tensor field  theories
for which the gauge symmetry is $U(1)$ rather than the real line 
\cite{polyakov}.
Notice that, as in the 2D case, the coupling constant $g$ could have been
 absorbed in the definition of the fields
entering then in the periodicity of the fields.
In that case we can see that the two dual variables are
related by a Poincar\'e duality:
%
%
$d\L=^*dB$.
And duality exchanges field equations with Bianchi 
identities.
For the particular case of 4D QED, this implies that 
duality exchanges  electric and magnetic fields which is the
original duality of Maxwell equations in vacuum mentioned
in the introduction.
Again, as in the 2D case, we can see $A_h$ as the gauge field
of the global symmetry and the Lagrange multiplier term as imposing
vanishing field strength.
Also to be more strict we should have introduced
external current couplings in the action so that the
path integral would be a function of that current
as in equation (5).
Duality then holds for any  correlation function.
Similar to the 2D case where the dilaton transformation was
imposed, if we consider topologically nontrivial 
spacetimes there would be an extra term proportional 
to the Euler number of the manifold \cite{barbon}, generalizing the
dilaton transformation in the 2D case. We will not consider further
this possibility.

Finally, as a gauge field naturally couples to a pointlike
charged particle by the coupling $\int A_\mu dx^\mu$,
the field $B_{h-1}$ naturally couples to an  $h-2$
dimensional object, a $p$-brane ($p=h-2$) which is  the
analogous of the electric charges in QED.  The dual
field $\tilde{B}_{d-h}\equiv \L_{D-h-1}$ couples to
a  magnetic $D-h-2$ dimensional brane
analogous to the magnetic monopole. We will then see in section
4 that the duality between electric charges and magnetic monopoles
is generalized to a duality between electric
$(h-2)$-branes and magnetic $(D-h-2)$-branes
\cite{khuri}.

\subsection{Abelian Bosonization}

Now that we have a definite prescription for duality we can 
apply it to any system with a global symmetry.
The simplest example that comes to mind is a free fermion  in 2D
with the Dirac action
\begin{equation}
S_F=  \int d^2x \;
\left( - \overline\psi
\dsl \psi +i J_\mu\overline\psi\gamma^\mu\psi \right)
\end{equation}
where we introduced explicitly the external source $J_\mu$ to 
keep track of the functional dependence of the partition 
function $Z[J]=\int{\cal D}\psi e^{iS_F}$.
Notice that this action has the global symmetry
$\psi\rightarrow e^{i\theta}\psi$, therefore we can just follow the
same prescription as above, {\it ie} we gauge the global symmetry
and impose the constraint of vanishing gauge field strength \cite{boson1}.
(The first order formalism used before does not
apply clearly in this case.)
We have then to consider the partition function
\begin{equation}
Z_G[ J] \equiv  \int \Scd \psi \; \Scd A_\mu \; \Scd \Lambda \;
e^{ i
 S_G(\psi,J_\mu+A_\mu,\Lambda)} 
\end{equation}
where
\begin{equation}
S_G\equiv
 \int d^2x\; \Biggl( \Scl_F(\psi,J_\mu+A_\mu) +
\hf \, \Lambda  \veps^{\mu\nu} F_{\mu\nu} \Biggr)
\end{equation}
 Since we have to fix the gauge symmetry,
it is convenient to work in the Lorentz gauge $\partial\cdot A=0$.
It is straightforward to see that integrating the Lagrange 
multiplier $\Lambda$ sets $F=0$ which together with the Lorentz gauge
allows us to set $A_\mu=0$, recovering the original action.
In   order to find the dual action we have to integrate first over
$\psi$ and $A_\mu$ and find the effective action $S_B[\Lambda,J]$.
This can be performed thanks to the well known calculation of 
Schwinger:
\begin{eqnarray}
& &\int \Scd\psi \; \exp\left[ i \int d^2x \Bigl( - \psibar \dsl \psi +
i \psibar
\gamma^\mu \psi \; A_\mu \Bigr) \right] \nonumber\\
&= &  \exp\left[
{i \over 4\pi} \int d^2x \;
F^{\mu\nu} \; \Square^{-1} \; F_{\mu\nu} \right].
\end{eqnarray}
Next we have to integrate over $A_\mu$. The Lorentz gauge
implies we can write $A_\mu=\veps_{\mu\nu}\partial^\nu\varphi$,
the corresponding jacobian is a constant and the integral over
$\varphi$ is gaussian. We end up with the bosonic lagrangian:
\begin{equation}
\Scl_B(\phi,J) = - \hf \; \partial_\mu \phi \, \partial^\mu \phi
 + \frac{1}{\sqrt{\pi}} \;
\veps^{\mu\nu} \,\partial_\mu \phi \, J_\nu 
\end{equation}
where we have defined $\phi\equiv \sqrt{\pi}\Lambda$.
Therefore we have shown that the actions for free fermions and free
bosons are dual to each other, reproducing the standard bosonization 
result in a fashion that makes bosonization and standard dualities
be recognized as two manifestations of the same effect.
By comparing the coefficients of the source $J_\mu$ in both actions
we can explicitly see the equivalence:
\begin{equation}
i \psibar \gamma^\mu \psi \leftrightarrow - \, \frac{1}{\sqrt{\pi}}
\;\veps^{\mu\nu}
\, \partial_\nu \phi
\end{equation}
which is the standard statement of bosonization. Coupling the axial  current  to
another source gives the mapping $i \psibar \gamma^\mu \gamma_3\psi \leftrightarrow 
 \, \frac{1}{\sqrt{\pi}}
\;
\, \partial^\nu \phi$.

Notice that a crucial step in our derivation of the bosonic action
was the use of the explicit calculation of the fermion determinant
by Schwinger.
For a more complicated fermionic theory, including four-fermion
couplings and mass terms there is no explicit calculation of the determinant
but we still can use an extra property 
of the system, this is that besides the vector-like phase symmetry we
used for dualization, there is also an axial symmetry of the fermionic
theory:$\psi\rightarrow \exp(i\alpha\gamma_3)\psi$. This symmetry is anomalous
but using the gauged action with the Lagrange multiplier constraint above
we can see that the $\L F$ term plays the role of a Wess-Zumino anomaly cancelling
term. This can be seen explicitly  by realizing that the anomaly can be 
canceled by asigning a transformation to $\L$ under axial transformation
$\L\rightarrow \L+\alpha/\pi$. This combined with the fact that 
the integration over the gauge field gives a delta function forcing the
identifications (23) in terms of expectation values,
 is enough to determine that the dual action is given
by (22), without the need of using the explicit fermion determinant.
A similar argument can be used to prove that the massive Thirring model
\begin{equation}
S=S_F+\int d^2x\left(\hf g^2(\psibar\gamma^\mu\psi)(\psibar\gamma_\mu\psi)
-m\psibar\psi\right)
\end{equation}
is dual to the sine-Gordon model
\begin{equation}
S_{SG}=S_B-Am \int d^2x\left(1-\cos{\beta\phi}\right).
\end{equation}
With $A$ an undetermined constant and $\beta$ is the coupling
given by $\beta=2\pi/\sqrt{\pi+g^2}$. We can then see that a 
weak coupling Thirring model corresponds to a strongly coupled
sine-Gordon model and viceversa.
This then reproduces  the well known results on bosonization.

The power of the present approach relies on the fact that 
it provides a unified way to treat bosonization and duality,
something that was not done in the past. Furthermore, it allows
to extend the concept of bosonization to any dimension $D\geq 2$
\cite{boson3}.
In an arbitrary dimension we can immediately see    that the
bosonic coordinate $\L$ will be extended from a scalar in 2D
to an antisymmetric tensor of rank $D-2$ in order to have
the Lagrange multiplier term in the Lagrangian $\veps_{M_1\cdots
M_D}F^{M_1M_2}\L^{M_3\cdots M_D}\equiv \L\cdot d A$.
\begin{equation}
\Scl_G = \Scl_F+m\psibar\psi + i \psibar \gamma^\mu \psi \, A_\mu +
\left(\L\cdot d  A\right)
\end{equation} 
 The problem we face in 
higher dimensions is that the fermion determinant is not 
computed, the symmetry argument we used does not help in 
higher dimensions mainly because of the properties of the $\gamma$
matrices. We should then rely on special cases. A simple
case is to consider a massive fermion in the limit of large
mass. For this case
\begin{eqnarray}
& &\int \Scd\psi \exp
\left\{ -i \int d^Dx \; \psibar \Bigl[ \gamma^\mu D_\mu + m \Bigr] \psi
 \right\}\nonumber \\
&=& \exp \left\{ {i \over 2} \int d^Dx \; A_\mu \Pi^{\mu \nu}_D
A_\nu + \cdots \right\}, 
\end{eqnarray}
 Where $D_\mu\equiv \left(
\partial_\mu - i A_\mu \right)$.
The vacuum polarization $\Pi^{\mu\nu}_D$ has been computed
in the past in a $1/m$ expansion, the first order term is:
\begin{eqnarray}
\Pi^{\mu\nu}_D&=&k_D\left(\Square\, \eta^{\mu\nu}-\del^\mu\del^\nu\right)\,
 ,\qquad
 D>3\\
&=&k_3\veps^{\mu\nu\lambda}\del_\lambda\qquad\qquad  D=3\nonumber
\end{eqnarray}
with $k_D$ a $D$-dependent constant and $k_3={\rm sign}\  m/8\pi^2$.
For $D>3$ this leads to the nonlocal action:
\begin{equation}
S=-\int\d^dx\left(\frac{1}{2k_D}\Omega_\mu
\frac{1}{\Square}\Omega^\mu+\Omega_\mu J^\mu\right).
\end{equation}
where $\Omega\equiv \tilde{\Lambda}$.
For $D=3$ however, we get a local action of the Chern-Simons type
\begin{equation}
S=-\int d^3x \veps^{\mu\nu\rho}\left(\frac{1}{2k_3}
\Lambda_\mu\del_\nu\Lambda_\rho+J_\mu\del_\nu\Lambda_\rho\, . \right)
\end{equation}
with the standard bosonization  relations
\begin{equation}
i \psibar \gamma^\mu \psi \leftrightarrow - \,
\;\veps^{\mu\nu\rho}
\, \partial_\nu \L_\rho\ .
\end{equation}

\section{NON--ABELIAN DUALITY }
In this section we will try to generalize the results of the
previous section for cases when the global symmetry is non-abelian.

\subsection{Non-Abelian $T$-Duality}

Consider  the $\sigma$--model action (1) and
assume that the target space metric has a group $\cal{G}$ of
non--abelian isometries \cite{nonab}, \cite{givroc}, \cite{ricci}, \cite{alvarez2}.
  In this case, $Q_{MN}$ {\it does}
depend on $X^m$ and transforms accordingly
under $X^m\to g^m_{\ n} X^n\ , g\in \cal{G}$.
  We gauge the symmetry corresponding to a
subgroup $H\subseteq\cal{G}$
    $\del X^m \to DX^m = \del X^m +
              A^{\alpha}(T_{\alpha})^m_{\ n} X^n\ $,
and add to the action the term
$\int d^2 z\ tr(\L F) =
                        \int d^2 z\ \L_{\alpha}F^{\alpha} ,$ 
where in this case the gauge field strength is,
in matrix notation,
$F =
       \del \overline A - \delbar A + [A,\overline A] $.
The  $N \times N$ matrices $T_{\alpha}$ form an
adjoint representation of the group $H$.
In the path integral we have to consider
\begin{eqnarray}
{\cal Z}&= & \int {{\cal D}X\over V_{\cal G}}\
       \int D\L\ DA\ D\overline A\ \\
    & &   e^{\left\{ -i\left( S_{G}[X,A,\overline A]
 +
             \int d^2 z\ tr(\L F) \right) \right\}} \ ,\nonumber
\end{eqnarray}
where $V_{\cal G}$ is the ``volume'' of the group of isometries
and ${\cal D}X$ is the measure that gives the correct volume
element
${\cal D}X = DX \sqrt{G} e^{-\Phi}\ $.
Similar to the abelian case, the original action is obtained by
integrating out the Lagrange multiplier $\L$.  Locally, this forces
the gauge field to be pure gauge
$A = h^{-1}\del h\ ,
    \ \overline A = h^{-1}\delbar h,\ h\in H $.
By fixing the gauge with the choice
$A = 0$, $\overline A = 0$ we reproduce
the original theory.  
The dual theory is obtained by integrating out
the gauge fields in the path integral (32). 
Integrating over the gauge fields $A$, $\overline A$ we obtain
\begin{equation}
               \int {\cal D}X\ D\L\
         \delta[{\cal F}]\ \det{\frac{\delta{\cal F}}{\delta\omega}}\
                              e^{-iS'[X,\L]} \det (f^{-1})
\end{equation}
 where ${\cal F}$ is the gauge fixing function and $\omega$ are the
parameters of the group of isometries.  The matrix $f$ is the 
coefficient of the quadratic term in the
gauge fields  and $\tilde{S}$ is given by
\begin{equation}
{\tilde{S}[X,\L] = S[X]
               - \frac{1}{4\pi\alpha'}\int d^2 z\
            \overline{h}_{\alpha} (f^{-1})^{\alpha\beta} h_{\beta}}
\end{equation}
Here  $h$ and $\overline{h}$
are the currents coupled to $\overline A$ and $A$ respectively 
\cite{nonab}.
After the gauge fixing,
denoting the new coordinates in the dual manifold collectively 
by $Y$,
we have
\begin{equation}
{\cal P} =
        \int {\cal D}Y\, e^{-i\tilde{S}[Y]} \det (f(Y)^{-1}) \ .
\end{equation}
The Fadeev--Popov determinant
in the path integral contributes to the measure such that
the correct volume element for the dual manifold is 
obtained
${\cal D}Y = DY \sqrt{\tilde{G}} e^{-\Phi'}\ $.
The factor $\det (f^{-1})$ in the partition function
can be computed using standard 
regularization techniques \cite{buscher}. It generates
a new local term in the action of the form
$\frac{1}{8\pi\alpha'}\int d^2 z\
         \alpha' R^{(2)}\ (\Delta\Phi ) $ ,
which corresponds to the change in the dilaton
due to the duality transformation
\begin{equation}
{\tilde{\Phi} = \Phi - \log\det f\ .}
\end{equation}
This change in the dilaton transformation is the
shift necessary to retain the conformal invariance
of the dual theory.

In general, we cannot write explicitly the gauge fixed
dual action. Therefore, we are not able to
present the new metric and antisymmetric tensor fields in a closed
form, as was done for the abelian case in equations (3).
As an example, let us
consider a theory  for which the target space metric
has a maximally symmetric subspace
with ${\cal G} = SO(3)$ and no
antisymmetric tensor. The coordinates $X^M, M = 1,...,D$,
can be decomposed into the two angular coordinates
($\theta$, $\phi$) describing
$2$--dimensional spheres, and $D-2$ extra
coordinates ($v^\mu$) specifying the different spheres in
the $D$ dimensional spacetime. The metric can then be
decomposed  in the form
\begin{equation}
S = S[v] + \int d^2 z\ a^2\Omega
  \left(\del\theta \delbar\theta +
          \sin^2\theta \del\varphi\delbar\varphi
                        \right)
\end{equation}
It is more convenient to
treat the coordinates $\theta$  and $\phi$ in terms
of cartesian coordinates $X^m$ in
3--dimensional space on which $SO(3)$ can
act linearly, so
we write the $\sigma$-model action in the form
%
\begin{eqnarray}
S &= &S[v]+\int d^2 z\ \Omega(v)
      \left\{\frac{}{} G_{mn}\del X^m \delbar X^n +\right.\nonumber \\
& & \left. \frac{1}{2a\sqrt{\Omega}}\lambda (G_{mn} X^m X^n - a^2)\right\}
 ,
\end{eqnarray}
where
$S[v] = \int d^2 z G_{\mu\nu}(v)\del v^{\mu} \delbar v^{\nu}$,
the metric $G_{mn}$ is
diagonal and constant and
the Lagrange multiplier term fixes
the 3 dimensional space to be
a sphere of radius $a$.  
Gauging this action and fixing the gauge $A=\bar A=0$ we obtain
the original action.
Another  convenient choice of gauge is to set
$X^1=X^2 = 0, \quad X^3 = a$, $A^1 = \del\theta$,
$A^2 = - \sin\theta\del\varphi$ and
$A^3 = \cos\theta\del\varphi$ leading to the original action 
in 
spherical coordinates .
%

We can write a general expression for the dual action after fixing 
the coordinates
$X^m$ as above, but before fixing the remaining degree of freedom 
corresponding to one of the Lagrange multipliers $\Lambda_\alpha$.
We then have (from here on, we ommit 
writing explicitly the dilaton term of equation (1))
%
\begin{equation}
\tilde{S}= S[v]+\frac{1}{4\pi\alpha'}
                 \int d^2 z\,  \left(
               \del\L_{\alpha} (f^{-1})^{\alpha\beta} \delbar\L_{\beta}
                   \right)
\end{equation}
{}From this expression,  we can in principle read off
the new background fields
as in (3). This is actually the general expression for any group.
 But we  still  have to complete
the gauge fixing for the $\L_{\alpha}$. In our case,
 choosing $\L_2 = 0$ and defining
$ x^2 = \L_1^{\ 2} + \L_3^{\ 2}$ and $y = \L_3$, we obtain
the dual theory action
%
\begin{eqnarray}
    \tilde{S} =  S[v]+ \int d^2 z\
          \frac{ \left( a^4 \Omega(v)^2\del y \delbar y
                    + x^2 \del x \delbar x \right)}{
4\pi\alpha'a^2\Omega(v) (x^2 - y^2)}
\end{eqnarray}
and the dual dilaton field is now given by 
$\tilde{\Phi}  = \Phi - \log [a^2 \Omega(v)\ (x^2 - y^2)] $.

 Consider the dual geometry of Schwarschild with 
respect to the
$SO(3)$ symmetry.  We find
\begin{eqnarray}
ds^2&=&-(1-{2M/r}) dt^2 +\frac{dr^2}{ 1-{2M/r}}\\
  & & +\frac{1}{r^2 (x^2 - y^2)}\ [r^4 dy^2
                    + x^2 dx^2]\  ,\nonumber 
\end{eqnarray}
with the new dilaton $\tilde{\Phi}= \Phi - \log [r^2 (x^2 - y^2)]$. The
regions $x=y$
and $r=0$ are real singularities whereas
$r=2M$ is only a metric singularity
corresponding to a horizon as in the original case. 
Notice that the metric  is {\it not}
spherically symmetric,
in fact it does not have non-abelian isometries,
raising the question of how to dualize back,
this is a definite difference with the abelian case
and shows in some way that the global symmetry is not an 
essential ingredient to obtain duality. We will see that 
Poisson-Lie $T$-duality takes care of this situation
in an elegant way.
Again, it is straightforward to check that this background fields
satisfy the string background field equations
thus providing `new' string vacua.

To   finish this subsection we can mention some general features
of non-abelian duality. 
If the non-abelian group ${\cal G}$ is non-compact, there may
be an anomaly that prevents the dual model from being 
conformal. An example of this type was considered in
\cite{ricci} and the understanding of the anomaly was found
 in
\cite{alvarez2}. In this paper it is also realized that a better
treatment of non-abelian duality is to express the dual
theory in terms of a group-valued coordinate $g$ instead of
the algebra-valued coordinate $\L$, this can be achieved 
by performing a non-local change of variables
$\del \L\rightarrow g^{-1}\del g$. this allows to formulate 
non-abelian duality in a way closer to the abelian case,
it is also crucial for the non-abelian bosonization we will discuss
next. Furthermore, it has also been shown that non-abelian
duality can  also be realized as a particular canonical transformation
\cite{yolanda}.
Global issues are less under control for the non-abelian case
and not much progress has been achieved in this direction.
Finally it is worth remarking that a higher dimensional 
generalization of non-abelian duality can be thought in two 
ways, one is work with scalar fields in higher dimensions
defining a $\sigma$-model invariant under a  non-abelian
group. The steps to dualize are identical to what we have done
and again the complication is {\em only} technical. On the other
hand we could consider antisymmetric tensors of higher rank, 
like nonabelian gauge fields. A duality transformation has 
been attempted for this case which could be  relevant for 4D QCD for instance.
However, 
naive dualization takes a Yang-Mills gauge theory into another
theory which is not Yang-Mills
\cite{freedman}. Non-abelian 
antisymmetric tensor theories  of rank greater than two
have been  argued not to exist \cite{claudio}.

\subsection{Poisson-Lie T-Duality}
 A very elegant generalization of  non-abelian $T$-duality
is the Poisson-Lie $T$-duality introduced by Klim\v c\'{\i}k and \v Severa
\cite{klimcik}.
We will briefly explain here the main idea of this generalization.

If the $\sigma$-model (1) is invariant under the action of a group
${\cal G}$, the corresponding `Noether' current:
\begin{equation}
J_a=v_a^M\, Q_{MN}\left(\del X^M \, dz-\delbar X^N\, d\overline{z}\right)
\end{equation}
is conserved {\it i.e.} $dJ_a=0$. Here $v_a^M$ is the vector 
field associated to the right action of $G$. If the current is
 not conserved but satisfy:
\begin{equation}
dJ_a=\hf \tilde{c}_a^{kl}J_k\wedge J_l
\end{equation}
where $\tilde{c}_a^{kl}$ are the structure constants of another, dual
Lie group $\tilde{\cal G}$, then the $\sigma$-model is said to have 
the ${\cal G}$ 
Poisson-Lie symmetry with respect to $\tilde{\cal G}$.
Then we can generalize the idea of duality based on a global
symmetry to cases with Poisson-Lie symmetry.
In that case
we can  immediately see that abelian and nonabelian dualities are
just particular cases of this duality  where the dual group 
$\tilde{\cal G}$
 is abelian
and so the structure constants vanish. In that case the nature of 
${\cal G}$ 
distinguishes between abelian and non-abelian dualities.
This explains the lack of non-abelian isometries in the models
considered in the previous subsection. For non-abelian
$\tilde{\cal G}$ this is a new duality symmetry. Therefore
Poisson-Lie T-duality is a more general setting to formulate $T$-duality.
Duality here is manifest since the role of the two dual groups
${\cal G}$, $\tilde{\cal G}$ can be interchanged. These groups form what is
called a Drinfeld double, defined by the decomposition of a bigger group
 ${\cal D}$
into ${\cal D}={\cal G}+\tilde{\cal G}$
 where the algebras of ${\cal G}$, $\tilde{\cal G}$
 are maximally
 isotropic subalgebras of that of ${\cal D}$ with respect to a
 nondegenerate
 invariant bilinear
form of the algebra of ${\cal D}$.
Using this approach it has been possible to generalize some of the properties
of abelian  duality such as the momentum/winding exchange of the two dual 
models.
 
\subsection{Non-abelian Bosonization}

In the same way  that abelian bosonizaton was understood 
in section 2.3 as a particular case of
abelian duality, we can also understand non-abelian bosonization
as a particular case of  non-abelian duality
\cite{boson2}.
The starting point is a system of Majorana fermions in $1+1$
dimensions, invariant under the global symmetry $O(N)$.
The partition function is
\begin{equation}
{\cal Z}=\int {\cal D}\psi\exp{\left(-\frac{i}{2}\int d^2 x\, 
\psibar\gamma^\mu\left(\del_\mu-a_\mu\right)\psi\right)}
\end{equation}
with $a_\mu$ being a  matrix valued external fields.
Again dualization can be ahieved by gauging the anomaly 
free $O(N)$ group and imposing the corresponding field strength to
vanish.
\begin{equation}
{\cal Z}=\int {\cal D}\psi{\cal D}A{\cal D}\L
e^{i\int d^2 x\left(-\hf \psibar\gamma^\mu D_\mu\psi
+\veps^{\mu\nu}\L F_{\mu\nu}\right)}
\end{equation}
with the covariant derivative $D_\mu\equiv
\del_\mu-i(a_\mu+A_\mu)$. Fixing the gauge and performing 
the fermion integral will lead  a theory in terms
of $\L$ which is the desired bosonization.
Luckily, as in the abelian case, the fermion integral
has been already computed by Polyakov and Wiegmann
\cite{polwieg}. The result is better expressed by redefining
$A_\mu$ and $a_\mu$ in terms of group valued 
variables ($a_{\pm}=ih_{\pm}^\dagger\del_{\pm}h_{\pm}$ etc.).
\begin{equation}
{\cal Z}[a]={\cal Z}[a=0]\exp{\left(-i\Gamma[h_{-}h_{+}^\dagger]\right) }
\end{equation}
With $\Gamma$ the Wess-Zumino-Witten action \cite{polwieg,witten1}.
Using this and performing a nonlocal change of variables for
the algebra-valued Lagrange multiplier $\del\L\leftrightarrow
g^\dagger\del g$, we find that the partition function
is \cite{boson2}
\begin{equation}
{\cal Z}_G[a]= \int {\cal D}g\exp{i\Gamma_{GWZW}[g,a]}
\end{equation}
With $\Gamma_{GWZW}$ the gauged Wess-Zumino-Witten
action $\Gamma_{GWZW}[g,a] \equiv  \Gamma[h_{-}gh_{+}^\dagger]
-\Gamma[h_{-}h_{+}^\dagger]$. We have then seen that the bosonized
action is the level one WZW action dual to the action for free Majorana
fermions. By comparing the couplings to the external field $a_\mu$
we can also get the standard bosonization relations
\begin{equation}
i\psibar\gamma_-\psi\leftrightarrow
\frac{i}{2\pi}g^\dagger\del_- g ,
\quad i\psibar\gamma_+\psi\leftrightarrow-\frac{i}{2\pi}
\del_+gg^\dagger
\end{equation}

\section{BROKEN GLOBAL SYMMETRIES}

In this section we will concentrate on the general antisymmetric tensor field
theories discussed in section 2 and abelian duality. The question we
will pose is if the existence of the global symmetry is not only 
a sufficient but also necessary condition for duality. The answer is no.
A way to illustrate this is by starting with  a system with a
global symmetry and break it by the existence of topological defects.
Then we can ask what happens to duality. We will see that duality
not only survives this breaking but gets enriched in the sense
that it relates different phases of the theory.

\subsection{QED and monopoles}

Let us start with the best known case of duality which is
Maxwell's equations. We know that in vacuum these equations are
invariant under the exchange of electric and magnetic fields.
Duality exchanges the two Bianchi identities with the two 
field equations by exchanging the field strength $F_{\mu\nu}$
with its dual $\tilde{F}_{\mu\nu}\equiv \veps_{\mu\nu\rho\sigma}
F^{\rho\sigma}$. Duality was easily derived in section 2.
However this was true only for the case of Maxwell's equations
without sources. In order to generalize this duality to the case
with sources we have to follow Dirac and include electric and magnetic
sources. The simplest case is the source due to the presence
of a   `elementary' electric charge $q_e$
and a `solitonic'  magnetic charge $q_m$ for which
the corresponding currents are:
\begin{equation}
J_{e,m}^\mu=q_{e,m}\int d\xi^\mu\delta^4(x-\xi).
\end{equation}
In order to derive duality for this case we first need to write an
action that incorporates both sources. Inclusion of the elctric
source is straightforward, we only need to add a term of the
form $A_\mu J_e^\mu$ to the Maxwell action. The magnetic current
is not that simple but it can be done by first noticing that in 
the presence of a magnetic monopole the field strength is no longer 
of the form $F_{\mu\nu}=\del_\mu A\nu-\del_\nu A_\mu$.
 In fact the 
magnetic field due to a monopole can be written as
\begin{equation}
\vec{B}=\frac{q_m\hat{r}}{4\pi r^2}=\nabla\times \vec{A}-
q_m\, \theta(-z)\, \delta(x)\, \delta(y)\, \hat{z}
\end{equation}
where the last term represents the Dirac string along the
$-z$ axis. Writing this in a  covariant form implies
\begin{equation}
F_{\mu\nu}=\left(\del_\mu A_\nu-\del_nu A_\mu\right) -V_{\mu\nu}
\end{equation}
with $V_{\mu\nu}\equiv q_m \veps_{\mu\nu\rho\sigma}\int\delta^4(x-\xi)
d\xi^\rho\, d\xi^\sigma$,
from which we can easily see that
 $\del_\nu\veps^{\nu\mu\rho\sigma}{V}_{\rho\sigma}=
J_m^\mu$.
 Therefore we can write the action
as:
\begin{equation}
S=\int d^4x\left( F_{\mu\nu}^2+A_\mu J_e^\mu\right)
\end{equation}
with $F_{\mu\nu}$ as given above. The first observation we can make 
is that the field equations derived from this action are precisely
Maxwell's equations in the presence of the electric and magnetic sources:
\begin{equation}
\del_\nu\, F^{\mu\nu}= J_e^\mu\ \qquad , \,\qquad \del_\nu\tilde{F}^{\mu\nu}=J_m^\mu.
\end{equation}
Duality is not trivial for this action since tere is no  longer 
the global symmetry shifting the field $A_\mu$.
Nevertheless, knowing how $F_{\mu\nu}$ differs from
a total derivative allows us to still be able to dualize this system.
In order to find the dual action,  we start with the first order
Lagrangian (with $F_{\mu\nu}$ arbitrary)
\begin{equation}
{\cal L}=\left(\left(F_{\mu\nu}-V_{\mu\nu}\right)^2
+\tilde{F}_{\rho\sigma}\left(\del_\rho\L_\sigma-
\tilde{V}_{\rho\sigma}\right)\right)
\end{equation}
Where $\tilde{V}_{\rho\sigma}$ satisfies $\del_\nu\tilde{V}^{\mu\nu}=J_e^\mu$. 
{}From this we can see that integrating out $\L_\sigma$ implies 
$F_{\mu\nu}=\del_\mu A_\nu-\del_\nu A_\mu$ and we get back the original action.
Changing the order of integration, we can integrate $F_{\mu\nu}$
and get the dual Lagrangian with $\L$ as the fundamental variable:
\begin{equation}
\tilde{\cal L}=\left(\left(\del_\mu\L_\nu-\del_\nu
\L_\mu-\tilde{V}_{\mu\nu}\right)^2-
\L_\mu J_m^\mu\right)
\end{equation}
where we have used that $\int d^4x\veps^{\mu\nu\rho\sigma}V_{\mu\nu}
\tilde{V}_{\rho\sigma}=q_e q_m I=2\pi n$  where $I$ is the integer
intersection number of the surfaces defined by $V_{\mu\nu}$ and
$\tilde{V}_{\rho\sigma}$ and we have used the Dirac quantization condition
$q_e q_m=2\pi$ integer.
{}From this pair of dual  actions  we can explicitly see the realization of
 electric/magnetic duality 
even in the presence of monopole sources that break the global symmetry.
A generalization to any  dimension and arbitrary antisymmetric tensors
is straightforward.

\subsection{Monopole Condensation}

We saw in the previous subsection how duality survived in the 
presence of electric and magnetic monopoles. Now we can imagine
what happens when we have a continuous distribution of magnetic
monopoles, {\em i.e.} monopole condensation.
In this case the field $F_{\mu\nu}$ is not the derivative of
$A_\mu$ not only at the Dirac string of a single monopole but 
for all spacetime points. Therefore in this situation it is natural
to assume $F_{\mu\nu}$ as the fundamental field and write an effective 
action for it. Imposing that this action be local, Poincar\'e invariant,
with no more than  two derivatives
and that it reduces to the Maxwell action in the limit of vanishing
density of monopoles $\rho$, it uniquely fixes the action to be:
\begin{equation}
S_{conf.}=\int d^4x\left(-\frac{1}{e^2}F_{\mu\nu}^2+\frac{1}{\rho^2}
(\del_\mu F_{\nu\alpha})^2\right)
\end{equation}
we can easily see that if the monopole density vanishes only the
configurations for which $(\del_\mu F_{\nu\alpha})=0$
contribute to the path integral implying $F=dA$ and reducing
to the Maxwell action. This action describes a massive 
2-index antisymmetric tensor field with mass $M=\rho/e$.
We have added the   subscript {\it conf} to the action since we will
see it will describe the confining phase of QED.

At this point it is natural to ask whether  this action has a dual
and if there is a way to find it. Notice that at this level there
are no traces of global symmetries. Nevertheless we can start with the
first order action:
\begin{equation}
S=\int d^4 x\left(
\frac{-1}{e^2}F_{\mu\nu}^2-\veps^{\mu\nu\rho\sigma}\tilde{B}_\mu\del_\nu
F_{\rho\sigma}-\rho^2\tilde{B}_\mu^2\right)
\end{equation}
There is no Lagrange multiplier appearing linearly in this
action, but the fields $\tilde{B}_\mu$ and $F_{\mu\nu}$ 
appear quadratically and therefore we can still integrate them out 
by means of a gaussian integration. Integrating out $\tilde{B}_\mu$ 
implies $\rho^2 \tilde{B}_\mu=\veps_{\mu\nu\rho\sigma}\del^\nu F^{\rho\sigma}$
and substituting back into the action reproduces the original theory $S_{conf}$.
Integrating out $F_{\mu\nu}$ instead, leads to the dual action:
\begin{equation}
\tilde{S}_{conf}=\int d^4 x\left(-e^2\left(\del_\nu\tilde{B}_\mu\right)^2
+\rho^2\, \left(\tilde{B}_\mu\right)^2\right)
\end{equation}
This action describes a massive vector having one propagating 
degree of freedom with mass $M=\rho/e$. Therefore we can see that
duality survives the breaking of the global symmetry, but now
is a completely different duality. Instead of relating two massless
vectors as in the original (Coulomb) phase, it is now a duality
between a massive 2-index field and a massive vector, both describing the
confining phase of compact QED after monopole condensation.
Since in the Coulomb action, duality there was a symmetry between electric
and magnetic degrees of freedom,
 we can ask if starting from the dual Coulomb action,
we could consider the condensation of electric monopoles giving rise not
to a confining phase but to a `Higgs' phase. The process is identical to the
confining case because the two dual actions for the Coulomb phase 
were the same (except for the change of $e\leftrightarrow 2/e$).
Therefore we get two dual actions for the Higgs phase also describing
three  massive degrees of freedom in terms of a massive vector
or massive two-form, with mass $\tilde{M}=\tilde{\rho}\, e$.
We can then see that the Higgs and confining phases are identical in this
case after exchanging $\tilde{\rho}\leftrightarrow\rho$ and
$e\leftrightarrow 2/e$. This  is known as Higgs/confinement
duality.

\begin{table*}[hbt]
\setlength{\tabcolsep}{1.5pc}
\newlength{\digitwidth} \settowidth{\digitwidth}{\rm 0}
\catcode`?=\active \def?{\kern\digitwidth}
\caption{Duality for the  Coulomb, Higgs and confinement phases}
\label{tab:effluents}
\begin{tabular*}{\textwidth}{@{}l@{\extracolsep{\fill}}rrrr}
\hline
\\
                 & \multicolumn{1}{l}{Original Lagrangian} 
                 & \multicolumn{1}{l}{Dual Lagrangian} \\

\\
\hline
\\
Coulomb phase    & $  \frac{1}{e^2}\left(dB_{h-1}\right)^2$ &  $ 
 \frac{e^2}{4}\left(d\tilde{B}_{d-h}\right)^2$ \\
\\
\hline
\\
Confinement phase & $ \frac{1}{\rho^2}\left(dH_h\right)^2+
\frac{1}{e^2}\left(H_h\right)^2
 $ &          $  \frac{e^2}{4}
\left(d\tilde{B}_{d-h}
\right)^2+\frac{\rho^2}{4} \left(\tilde{B}_{d-h}
\right)^2 $ \\
\\
\hline
\\
Higgs phase     & $\frac{1}{\tilde{\rho}^2}\left(
d\tilde{H}_{D-h}\right)^2+e^2\left(\tilde{H}_{D-h}\right)^2$ &$
\frac{4}{e^2}\left(dB_{h-1}\right)^2+\frac{\tilde{\rho}^2}{4}
\left(B_{h-1}\right)^2$\\
 \\

\hline

\end{tabular*}
\end{table*}

\subsection{Generalizations}

Now we can generalize these results to any dimension
as in section (2.2)\cite{orland}, \cite{carlome}.
 In $D$ dimensions we know the Coulomb phase
is described  by  two dual actions in terms of
antisymmetric tensors 
$B_{h-1}$ and $\L_{D-h-1}$ respectively, both carrying 
$ \left( {D-2 \atop h-1} \right)$             
          degrees of freedom. In that case the 
`elementary' charges are $(h-2)$-branes whereas the 
`solitonic' or magnetic charges 
are $(D-h-2)$-branes (as we have seen from other 
lectures, a monopole is a $0$-brane, an instanton a $(-1)$-brane
etc.) which are the natural objects coupling to  $B_{h-1}$ and $\L_{D-h-1}$
respectively. The confining phase is defined by the condensation  of 
the magnetic objects and is described by a massive antisymmetric
tensor $H_h$ dual to another  massive antisymmetric tensor
$\tilde{B}_{D-h-1}$ carrying $\left( {D-1\atop h}
\right)$   degrees of freedom with mass
$\rho/e$.
The Higgs phase can be generated by the condensation of the elementary
$(h-2)$-branes and would be described by massive antisymmetric
tensors $\tilde{H}_{D-h}$ dual to $B_{h-1}$, both carrying 
$D-1/h-1$ massive degrees of freedom 
of mass $\tilde{M}=\tilde{\rho}e$.
The effective Lagrangians for each of the three pairs of dual actions
are  shown in the table \footnote{The relative signs of the terms are dimension
dependent and are not shown here, see reference \cite{carlome}\
for the explicit signs on each case.}.

Some observations are in order here.
First, only for the case $h=(D-1)/2$ can we have Higgs/confinement
duality in which case it is realized as the 4D QED case.
Notice that the dual Higgs phase description is natural in the sense
that it corresponds to  the original field $B_{h-1}$ getting a mass
as in the original Higgs effect, this analogy can be seen more explicitly 
by realizing this effect in terms of a Stuckelberg mechanism.
This is similar to the dual confing phase in which the original dual field
$\tilde{B}_{d-h}$ gets a mass. We can then conclude that the  mechanism
to describe the confinement and Higgs phases by taking the
original field strength and promoting  it to the fundamental 
propagating field is just the dual of the Higgs effect.

I would like to remark  that the   process of monopole condensation is a dynamical
question that has not been addressed here. We have only assumed that this
happens and then described the effective actions and duality after condensation.
To know if the dynamics favours condensation we have to consider the
system in more detail and analyze  case by case.

The best studied examples are those in dimensions $D\leq 4$
with instantons or monopoles as the topological defects.
The 2D case  is particularly interesting since it corresponds
to strings on a circle as discussed in section (2.1)
for which the original and dual fields are 2D scalars $X,
\tilde X$. The elementary and topological defects are instantons 
($h-2=D-h-2=-1$).
The contribution of one type of instantons to the partition function
can be computed explicitly in the dilaton gas approximation
leading to an effective action (in the Higgs phase)
\begin{equation}
S=\int d^4 x\left( R^ 2\del X\delbar X-A\cos\frac{X}{R}\right)
\end{equation}
This is just the Sine-Gordon model discussed in section (2.3).
For small field values $X$ the potential reduces to a quadratic
potential
reproducing our result for the Higgs phase
(or dual confinement if we used $\tilde X$)(see table).
This is a check that our presciption to build the 
confinement effective action is correct (up to higher
derivative terms). The renormalization group flow for the
couplings $R,\rho$ in the Sine-Gordon model has been
well studied \cite{savit},  leading to the conclusion that at $R=2$
there is a phase transition from a Coulomb phase $R>2$
which is conformal invariant $\rho=0$, to a confinement 
or Kosterlitz-Thouless phase at $R<2$ which is not conformally
invariant and the field $X$ looses its interpretation as a 
coordinate in target space. 
The critical value $R=2$ can be found from the mass formula
(4), because for $R<2$ there appear tachyonic {\em winding}
 modes
\cite{sathiapalan}. The existence of these two phases has lead some
authors to conclude that instantons break the $R \leftrightarrow 
1/R$ duality. From our analysis we can see that this is not the case.
First, in the confinement phase duality is still manifest
but  now as a duality
between a massive scalar $X$ and a massive vector. Furthermore,
if we consider both types of instantons the phase structure 
is completely changed. This can also be seen from the  mass formula
(4) since also for $R>1/2$ there are extra tachyonic states,
but this time these are momentum states rather than winding.
The structure of the phase diagram happens to be identical to the
clock models studied by Cardy and  Rabinovici since they arrived at a 
condition which is {\em identical} to the existence of tachyons in 
equation (4). The  end result is that there are only two 
phases, Higgs and confinement (the original Coulomb phase is
not realized which is expected for bosonic strings because of the
 tachyon), separated at the selfdual point $R=1$, both phases are
described by massive scalars dual to massive vectors and so
the effective actions are identical except for the exchange
$R \leftrightarrow 1/R$ showing that the phase diagram is
manifestly invariant under duality! This is a realization of the 
Higgs/confinement duality.

This situation can be easily generalized for the 2D 
toroidal compactifications of section (2.1) but the phase structure
is much richer, the reason being the periodicity of the real
part of the field $T$ in (11) implying the existence of
an infinite number of oblique confinement phases separated at 
the points $T=\exp{\pi/3}$ and its infinite images under
$SL(2,\IZ)$ transformations \cite{cardy}.

For supersymmetric string theories there are no points 
where states become tachyonic and therefore only the Coulomb
(conformal) phase is realized \footnote{Remember this analysis
is non-perturbative in the 2D worldsheet but is still perturbative
in the string coupling, non-perturbative string effects can change 
this phase structure as we are learning from $S$ duality.}. But if 
the coordinate $X$ is time, then there are no tachyons for
temperatures smaller than the Hagedorn temperature $T_H$ where a
 tachyon appears indicating a phase transition, the tachyon dissapears
at the dual Hagedorn temperature $\tilde{T_H}$. Therefore there
are two Coulomb phases dual to  each other with the standard string 
picture and for temperatures $\tilde{T_H}>T>T_H$ there is a selfdual
phase described by a massive time coordinate. Again duality is
manifest.

For 3D QED again the topological defects are 
instantons and a dilute instanton gas  calculation
gives an action for the dual field (a scalar in this case)
of the Sine-Gordon type. It is well known that for this case
the dynamics is such that only the confining phase is realized.
 Our analysis only adds
that this phase can be described by two dual
actions in terms of a massive scalar or a massive 2-index
antisymmetric tensor.

For 4D QED a similar analysis can  be
done and the confining phase, induced by the condensation 
of monopoles,  can be described either
by  massive vector or a massive antisymmetric tensor with  two 
indices.  This case has been recently studied in detail by 
Polyakov who found that
the description in terms of a massive two-index tensor has the 
interpretation
of a `confining' string, since this field couples naturally to a string
\cite{polyakov2}, \cite{cristina}.
In this way understanding confinement by monopole (4D)
or instanton (3D) condensation, 
or by a confining string as in the old string 
picture of quark confinement can be
seen to be one and the same physical interpretation.
It has been claimed that this picture generalizes to Yang-Mills theories
\cite{polyakov2}.

Also for the D case with $h=3$, it has been seen explicitly that 
breaking of the global Peccei-Quinn symmetry
by non-perturbative effects,  does not 
break duality and the dual of the massive axion field is a massive
$3$-index field
\cite{bdqq}, in perfect agreement with our general results.

For higher dimensions and ranks of the tensors there are indications
that $p$ branes could also  condense and induce a phase transition.
This could be relevant for uncovering all the phases of
$M$ theory. However these cases need to be better understood,
 probably the recent studies of $D$-branes
may help understanding the  possibility of $p$-brane
condensation.

\subsection{Duality and Fourier transforms}

The duality between massive antisymmetric tensors
can be generalized for actions with higher derivatives and
general potentials. Starting with the action:
\begin{equation}
S=\int d^D x\left(\frac{}{} F(\del H_h)+ G(H_h)\right)
\end{equation}
This can be obtained from the partition function
\begin{equation}
\int {\cal D}\tilde{B}_{d-h}{\cal D} H_{h}e^{\int d^D x\left(H\cdot 
d\tilde{B}_{d-h} +G(H_h)+\tilde{F}(\tilde{B}_{d-h})\right)}
\end{equation}
where $e^{\tilde F}$ is the Fourier transform of $e^F$
\cite{kawai}.
Performing the integral over $\tilde{B}_{d-h}$ we recover the
original action above and performing the integral over
$H_h$ we obtain the dual action
\begin{equation}
\tilde S=\int d^D x\left( \tilde{F}(\tilde{B}_{d-h})+ \tilde{G}
(\del \tilde{B}_{d-h})\right)
\end{equation}
From here we can see that a nontrivial potential $G(H_h )$
gives rise to a higher derivative term for the dual $\tilde G(
\del \tilde{B}_{d-h})$ and viceversa. Also for a quadratic potential,
we know that the Fourier transform of a gaussian is another gaussian 
and so we recover the result of the massive actions of the
previous subsection. Furthermore, for vanishing potential
we know that the Fourier transform of $1$ is $\delta(\del 
\tilde{B}_{d-h)}$ and therefore we recover the duality of the 
Coulomb phases of section 2.2.
Therefore we can see that the dualities we have been dealing with 
for antisymmetric tensors are only particular cases of Fourier transforms
and finding the dual action reduces to finding Fourier transforms.
We can then consider a nontrivial example such as the sine-Gordon 
model which has a nontrivial
potential. In this case we know that the Fourier transform of 
$e^ {A\cos \tilde{B}_{d-h}}$ is the modified Bessel function $I_P(A)$
and therefore we can just read the dual higher-derivative action
 changing $P$ by $\del H_h$. This has been recently done for
QED ($h=2$) and the resulting (higher derivative) theory 
can give some information. First, since the original variable
$\tilde{B}_{d-h}$  is periodic, the Fourier transform is such that  the
summation in the path integral has to include  infinite
branches and this implies that in the confining string
picture we have to sum over all Riemman surfaces
\cite{polyakov2}\ making explicit
the string theory description of the confining phase.
Furthermore, the periodicity of the original variable
implies the quantization of flux for the dual
 variable. These features can not be seen using the quadratic
potential of the previous subsection where for instance periodicity
is not manifest.

It is curious to see that for the 2D case, this implies that
the Sine-Gordon model is dual to a (higher derivative) massive
vector model. But we had seen in section 2.3 that the Sine-Gordon
model is dual to the fermionic Thirring model, showing the richness of
duality!.

Notice that being the path integrals non-gaussian, most
of these results are only formal and a 
regularization technique has to be used in
order to make sense of the dual action.
For the $\cos$ potential mentioned above, a detailed lattice
calculation can be found in \cite{cristina}.

\section{OPEN QUESTIONS}

We have learned that global symmetries are a very useful tool
to prove the duality among different theories. In fact 
we can now take any physical system with a  global symmetry and dualize it.
For instance we could take the Schrodinger equation, obtained 
from a variational principle, and knowing that there is a phase
symmetry of the wave function, we can find the dual of 
Schrodinger's equation (the result does not look very illuminating at first 
sight).  However the 
existence of the global symmetry  is only a sufficient condition for duality
and it is not necessary. A more general statement can be  made
at least for the antisymmetric tensor field theories for which 
duality is only a functional Fourier transform. This may have 
interesting consecuences, in particular we may  
state an uncertainty principle regarding  dual variables analogous to
position/momenta in quantum mechanics.

An interesting field by definition has many open 
questions. I would like to mention a few.
\begin{enumerate}
\item {\it Superduality:}
We have used the existence of global `internal' symmetries as guidelines
for duality, but we know there are also global `spacetime' 
symmetries such as Poincar\'e and supersymmetry.
In this case we may start with a globally supersymmetric theory,
making   it local becomes supergravity and then we have to impose 
the constraint that the space is flat and the gravitino field
vanishes, again exchanging the order of  integration  implies
the existence of a dual theory, `superduality'. Work is
 in progress to present nontrivial
examples of this new duality \cite{super}.
\item  {\it Fermionization:}
We have seen how to go from a  purely fermionic theory to a bosonic
one, but the inverse process is not under control. For that we would
need to start with a bosonic theory and impose a constraint with a fermionic
Lagrange multiplier, which is not clear how to implement.
Probably the development of superduality may help in solving this
question since the constraint of imposing vanishing gravitino field
is fermionic.
\item {\it Mirror symmetry:}
Mirror symmetry is one of the most relevant symmetries discovered
 in perturbative string vacua. It sates that a
 compactification on a complex manifold
${\cal M}$ is equivalent to a compactification in a mirror manifold
${\cal W}$ such that the complex and K\"ahler structures of both manifolds
are interchanged. For Calabi-Yau spaces this corresponds to exchanging the
Hodge numbers $h_{11}\leftrightarrow h_{12}$.
Most of the evidence supporting mirror symmetry is `experimental',
in the sense that explicit construction of models show this symmetry
\cite{mirror}.
More recently, the toric  variety construction of Calabi-Yau manifolds
has allowed to assign ${\cal M}$ and ${\cal W}$ to two dual lattices
making mirror symmetry more explicit
\cite{mirror}. Also
 use of $T$-duality for different cycles of 
Calabi-Yau spaces has allowed to better understand mirror symmetry
\cite{strominger}.
     But an explicit proof of mirror symmetry is still lacking.
Previous attempts of proving mirror symmetry from $T$-duality have failed
\cite{pressel}.
Since Calabi-Yau manifolds have no isometries, the corresponding 
 $\sigma$-model action does not have the  global symmetries
usually  required to dualize.
In fact it was claimed in the past that the problem of proving mirror symmetry
was the same as proving duality without global symmetries. But this is
precisely what was done in chapter 4. However, without doing any calculation we can 
immediately see that the techniques of chapter 4 cannot work entirely to
prove mirror symmetry
because they would give for the dual of a $\sigma$-model with variables
$X^M$ (witha Landau-Ginzburg potential), a
 2D theory in terms not of coordinates $\tilde X^M$ but 
of `massive' vector variables $(V_\mu)^M$, which have no straightforward
identification with the coordinates of the mirror manifold. The duality
beween ${\cal M}$ and ${\cal W}$ seems closer to the Higgs/confinement 
duality of section 4.3 (see also \cite{witten3}\ ).
On the other hand the 2D $\sigma$-model does have a  global symmetry which 
is a $(0,2)$ supersymmetry on the worldsheet, therefore we may imagine that
superduality may have interesting relation with mirror symmetry.
 Notice that the $U(1)$ $R$-symmetry associated to
this $N=2$ supersymmetry  is also a global symmetry of the model, but
it is easy to see that dualizing with respect to this symmetry gives rise to
bosonization as in chapter 2 and not to mirror symmetry. 

\item {\it Seiberg's duality:}
Another interesting duality symmetry is the one proposed by Seiberg
for 4D $N=1$ supersymmetric field theories. This duality relates
different phases of an $SU(N_c)$ theory with those  of 
$SU(N_f-N_c)$ where $N_f,N_c$ represent the number of flavours and
colors respectively. Proposing this new duality symmetry 
 has been one of the major advances in the
nonperturbative understanding of supersymmetric field theories
recently. 
Again, the evidence for this duality is only indirect, although
it is very convincing.
An explicit proof following the lines of this paper is still
lacking. There are plenty of global symmetries in these systems
(flavour symmetries and supersymmetry), but, similar to the
case of higher dimensional bosonization, the highly nontrivial
nature of 4D field theories, combined with the
structure of the duality itself (see P. Argyre's 
lectures) makes an explicit proof of this duality
a very difficult and interesting challenge.

\item {\it Quantum Hall effect:}
As it was mentioned in the introduction, there are proposals for
a duality symmetry in the quantum Hall effect.
This duality would be with respect to the complex parameter
$\tau\equiv \sigma_{xx}+i\sigma_{xy}$ where $\sigma_{xx}$
and $\sigma_{xy}$ are the longitudinal and transverse conductivities
of the Hall effect. Since $\sigma_{xy}$ is quantized, we expect the
shift by an imaginary  integer to be a symmetry such as the axion 
field in string theory. The arguments in favour of an infinite
dimensional discrete group $\Gamma\subset  SL(2,\IZ)$ are many
which  have left to a proposal for a $\Gamma$ symmetric
phase diagram, but the
most striking fact is that,  contrary to the previous dualities, there
exists real experimental evidence for the existence of this duality.
Recent experimental results show explicitly that the transformation
$\sigma_{xx}\leftrightarrow 1/\sigma_{xx}$ relates two different 
phases of this system.
Again, an explicit proof of this duality is not available yet.
\end{enumerate}

We have seen duality is present in many different physical systems
and well defined techniques can be used to establish the equivalence
among the dual theories. 
Duality has also been observed experimentally, even though it is
in a condensed matter system, it gives us confidence that it
is actually present in other systems. As we have seen during this 
school, duality played an important role in the
solution of $N=2$ supersymmetric theories,
it is currently opening the way towards understanding
black holes and probably uncovering the fundamental theory of nature.
Furthermore we have seen that there are many 
open questions regarding  duality symmetries which makes the
subject healthy and alive.
\begin{center}
{\bf Acknowledgements}
\end{center}
I would like to thank the organizers for hospitality. C.P. Burgess
for comments on the manuscript and S. Theisen for encouragement.

\end{document}